\documentclass{jltp}
\usepackage{graphicx}
\title{Energy Spectrum of Vortex Tangle}

\def\Vec#1{\mbox{\boldmath $#1$}}
\author{Tsunehiko Araki, Makoto Tsubota and Sergey K. Nemirovskii$^*$}
\address{Department of Physics, Osaka City University, Osaka 558-8585, Japan\\
$^*$Institute of Thermophysics, Academy of Science, Novosibirsk 630090, Russia}
\runninghead{T.Araki, M.Tsubota and S.K.Nemirovskii}{Energy Spectrum of Vortex
Tangle}
\begin{document}
\maketitle
\begin{abstract}
The energy spectrum of superfluid turbulence in the absence of the normal fluid is
studied numerically. In order to discuss the statistical properties, we calculated
the energy spectra of the 3D velocity field induced by dilute and dense vortex
tangles respectively, whose dynamics is calculated by the Biot-Savart law. In the
case of a dense tangle, the slope of the energy spectrum is changed at $k=2\pi/l$,
where $l$ is the intervortex spacing. For $k>2\pi/l$, the energy spectrum has
$k^{-1}$ behavior in the same manner as the dilute vortex tangle, while otherwise
the slope of the energy spectrum deviates from $k^{-1}$ behavior. We compare the
behavior for $k<2\pi/l$ with the Kolmogorov law.

PACS numbers:  67.40.Db, 67.40.Vs.
\end{abstract}


\section{INTRODUCTION}
The particular attention is recently focused on the similarity between superfluid
turbulence and conventional turbulence\cite{Vinen,Stalp,Nore}. In conventional
turbulence, the energy spectrum $E(k)$ has the Kolmogorov form $k^{-5/3}$ for
wavenumbers $k$ in an inertial range. Such spectrum was observed indirectly by Stalp
{\it et al}. in superfluid $^4$He above 1.4K\cite{Stalp}, in spite of the fact that
the rotational flow in the superfluid component must take the form of discrete
quantized vortex lines, and that there can be no conventional viscous dissipation in
the superfluid component. This is understood by the idea that the superfluid and the
normal fluid are likely to be coupled together by the mutual friction and to behave
like a conventional fluid\cite{Vinen,Barenghi}. Since the density of the normal
fluid is negligible at mK temperatures, an important question now arises: even free
from the normal fluid, is the superfluid turbulence still similar to the
conventional turbulence or not?

Recently Nore {\it et al}. solved the 3D Gross-Pitaevskii equation and discussed
similarity to the Kolmogorov law\cite{Nore}. However this simulation includes such
complicated compressible effects as the radiation of sound from the vortex lines,
the interaction between vortex lines and sound, etc. In order to consider pure
superfluid turbulence in a simpler situation, we study the energy spectrum of the 3D
velocity field induced by the vortex tangle under the vortex filament model which
describes the dynamics of incompressible fluid.

\section{NUMERICAL CALCULATION}
For superfluid $^4$He, the vortex filament model is very useful, because the vortex
core radius $a \sim 10^{-8}$cm is microscopic and the circulation $\kappa =9.97
\times 10^{-4}$cm$^2$/sec is fixed by quantum constraint. Thus the dynamics of
vortices is calculated numerically by the Biot-Savart law which is described in our
previous paper\cite{Tsubota}. In our calculation, a vortex filament is represented
by a single string of points at a distance $\Delta \xi$ apart. When two vortices
approach within $\Delta \xi$, it is assumed that they are reconnected. The
computational sample is taken to be a cube of size $L=$1cm. This calculation assumes
the walls to be smooth and takes account of image vortices so that the boundary
condition may be satisfied.

We will introduce the energy spectrum of the velocity field induced by the vortex
filament. Using Parseval's theorem $\int d\Vec{k} |\hat{\Vec{v}} (\Vec{k})|^2 =
(2\pi)^{-3} \int d\Vec{r} |\Vec{v} (\Vec{r})|^2$, the kinetic energy can be written
as
\begin{equation}
E=\frac{\rho_{\rm s}}{2}\int d\Vec{r}|\Vec{v}_{\rm s}(\Vec{r})|^2 =\frac{\rho_{\rm
s}}{2} (2\pi)^{3}\int d\Vec{k}|\hat{\Vec{v}}(\Vec{k})|^2, \label{eq.1}
\end{equation}
where $\rho_{\rm s}$ is the superfluid density. By using the relation $
\hat{\Vec{v}}(\Vec{k}) = i \Vec{k} \times
 \hat{\Vec{\omega}}(\Vec{k}) /|\Vec{k}|^2$, we obtain the kinetic energy:
\begin{equation}
E=\frac{\rho_{\rm s}}{2}(2\pi)^3 \int
d\Vec{k}\frac{|\hat{\Vec{\omega}}(\Vec{k})|^2}{|\Vec{k}|^2}, \label{eq.3}
\end{equation}
where $\hat{\Vec{\omega}}(\Vec{k})$ is the Fourier component of the vorticity
$\Vec{\omega}(\Vec{r})=\Vec{\nabla}\times \Vec{v}(\Vec{r})$. In the vortex filament
model, the vorticity can be defined as $\Vec{\omega}(\Vec{r})=\kappa \int d\xi
\Vec{s}'(\xi)\delta (\Vec{s}(\xi)-\Vec{r})$, so that $\hat{\Vec{\omega}}(\Vec{k})$
is given by
\begin{equation}
\hat{\Vec{\omega}}(\Vec{k})=\frac{\kappa}{(2\pi)^3}\int d\xi e^{-i}\Vec{^{s}} ^{
(\xi)\cdot} \Vec{^k}\Vec{s}' (\xi). \label{eq.4}
\end{equation}
A vortex filament is represented by the parametric form $\Vec{s}=\Vec{s}(\xi,t)$,
where $\Vec{s}$ refers to a point on the filament, the prime denotes differentiation
with respect to the arc length $\xi$ and the integration is taken along the
filament. The energy spectrum $E(k)$ is defined as $E=\int_{0}^{\infty} dk E(k)$.
Finally we obtain the energy spectrum:
\begin{equation}
E(k)=\frac{\rho_{\rm s}\kappa^2}{2(2\pi)^3}\int \frac{d\Omega_k}{|\Vec{k}|^2}\int
\int d\xi_1 d\xi_2 \Vec{s}' (\xi_1)\cdot \Vec{s}'(\xi_2) e^{-i}\Vec{^k} ^{\cdot}
{^(} \Vec{^s} ^{(\xi_1)-}\Vec{^{s}} ^{(\xi_2))}, \label{eq.5}
\end{equation}
where $d\Omega_k$ denotes the volume element $k^2 \sin \theta_k d \theta_k d \phi_k$
in the spherical coordinate. The energy spectrum $E(k)$ is calculated for the vortex
configuration $\Vec{s}(\xi)$ obtained by the simulation of the dynamics.

\section{ENERGY SPECTRUM OF VORTEX TANGLE}
In order to make clear the statistical properties of superfluid turbulence, we
calculated the energy spectra of the 3D velocity field for dilute and dense vortex
tangles respectively.

First we discuss the energy spectrum for a dilute vortex tangle. This calculation of
the dynamics is made by the space resolution $\Delta \xi =4.58 \times 10^{-3}$ cm
and the time resolution $\Delta t =6.25 \times 10^{-5}$ sec. Figure 1(a) shows the
initial configuration of four vortices.
\begin{figure}[tbhp]
\begin{minipage}{1.0\linewidth}
\begin{center}
 \includegraphics[width=0.6\linewidth]{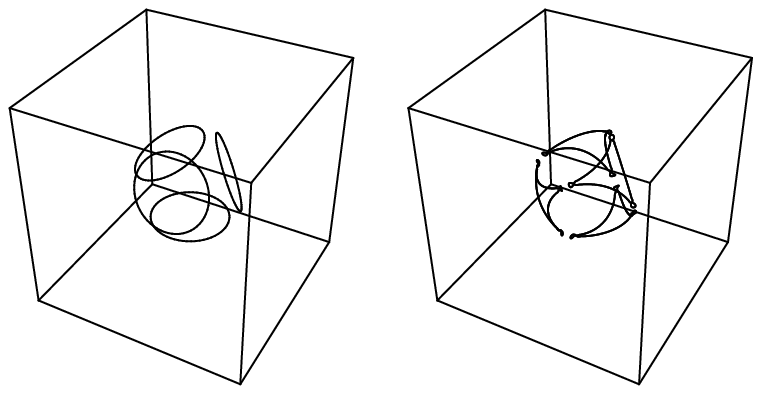}\\
 (a) \hspace{3cm} (b)
\end{center}
\end{minipage}
\\
\begin{minipage}{1.0\linewidth}
\begin{center}
 \includegraphics[width=0.6\linewidth]{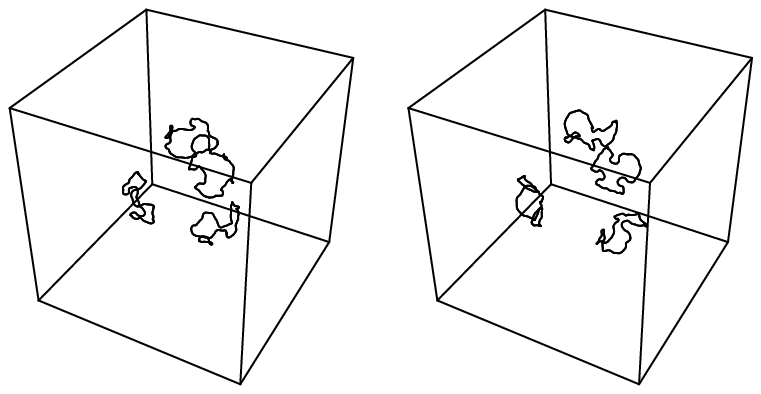}\\
 (c) \hspace{3cm} (d)
\end{center}
\end{minipage}
\caption{Collision of four rings at $t$=0 sec(a), $t$=6.25 sec(b), $t$=12.50 sec(c)
and $t$=18.75 sec(d). At $t$=6.12 sec, two vortices reconnect.}
 \label{eps1}
\end{figure}
 Four rings move toward the center of the cube by their self-induced
velocity to make the reconnection. After the reconnection the four rings move
outside oppositely(Fig. 1(b), (c), (d)). The time evolution of the energy spectrum
is shown in Fig. 2.
\begin{figure}[tbhp]
\begin{center}
 \includegraphics[width=0.6\linewidth]{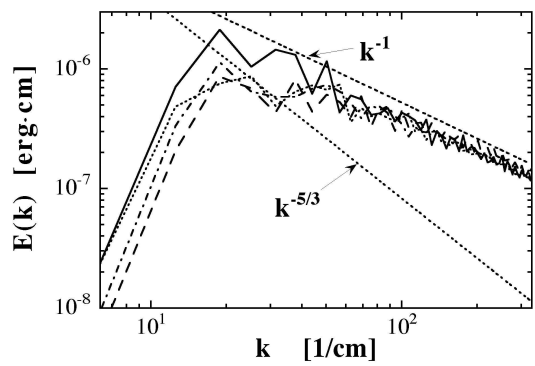}\\
\end{center}
\caption{Evolution of the energy spectrum of four vortices: $t$=0 sec(solid),
$t$=6.25 sec(long-dashed), $t$=12.50 sec(dot-dashed) and $t$=18.75 sec(dashed).}
 \label{eps2}
\end{figure}
 The slope of the spectra is not strongly affected by the vortex configuration.
 The energy spectrum $E(k)$ is proportional to $k^{-1}$, which comes from the
 velocity field of an isolated vortex\cite{Vinen,Kivotides}.

Next the energy spectrum for a dense vortex tangle is discussed. This calculation of
the dynamics is made by the resolution $\Delta \xi=1.83 \times 10^{-2}$ and $\Delta
t=1.0 \times 10^{-3}$sec.
 Figure 3(a) shows the initial configuration of the vortex tangle
\begin{figure}[tbhp]
\begin{minipage}{1.0\linewidth}
\begin{center}
 \includegraphics[width=0.6\linewidth]{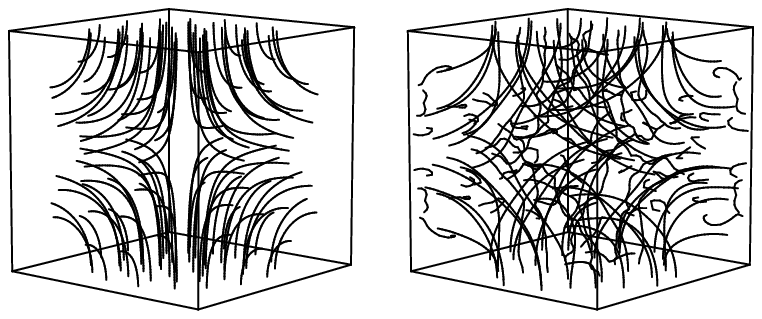}\\
 (a) \hspace{3cm} (b)
\end{center}
\end{minipage}
\\
\begin{minipage}{1.0\linewidth}
\begin{center}
 \includegraphics[width=0.6\linewidth]{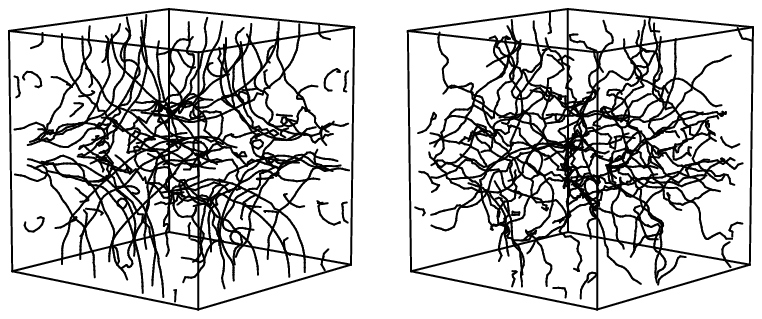}\\
 (c) \hspace{3cm} (d)
\end{center}
\end{minipage}
\caption{Time evolution of the vortex tangle at $t$=0 sec(a), $t$=30.0 sec(b),
$t$=50.0 sec(c) and $t$=73.0 sec(d).}
 \label{eps3}
\end{figure}
where the direction of the circulation of upper vortices is opposite to that of
   lower vortices(Taylor-Green vortex\cite{Nore}). The vortices become tangled through
    lots of reconnections(Fig. 3(b), (c) and (d)).
     The energy spectra are shown in Fig. 4.
\begin{figure}[tbhp]
\begin{minipage}{1.0\linewidth}
\begin{center}
 \includegraphics[width=0.8\linewidth]{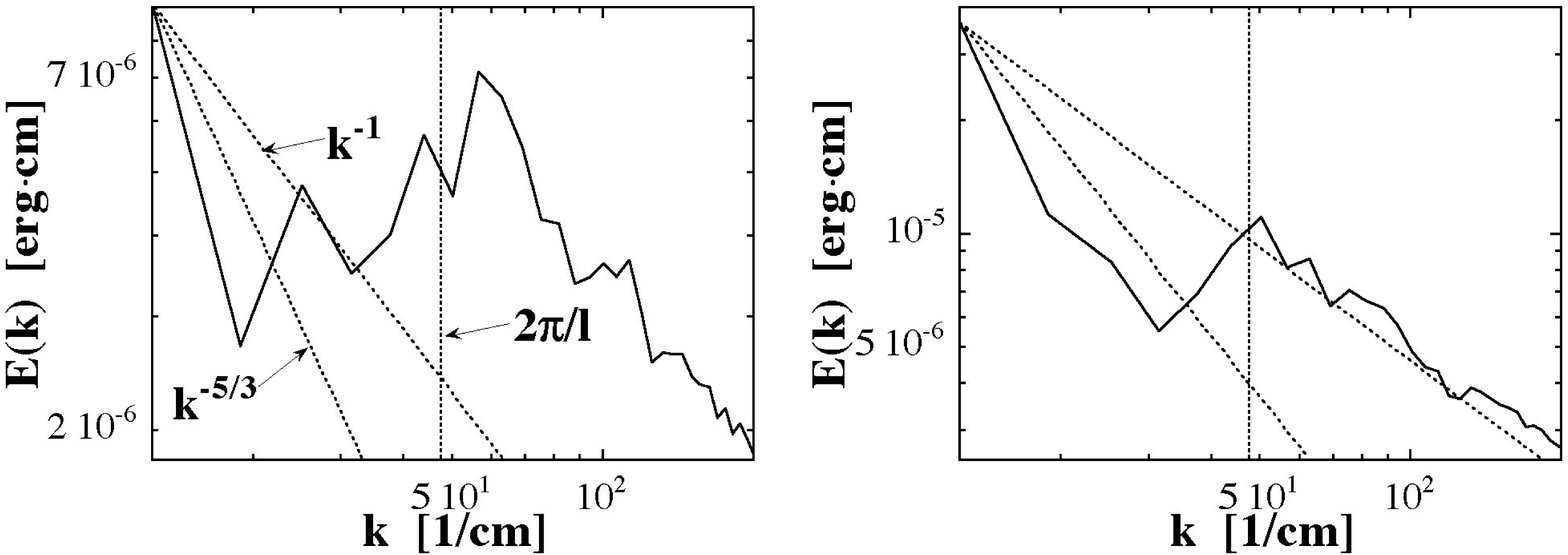}\\
 (a) \hspace{3cm} (b)
\end{center}
\end{minipage}
\\
\begin{minipage}{1.0\linewidth}
\begin{center}
 \includegraphics[width=0.8\linewidth]{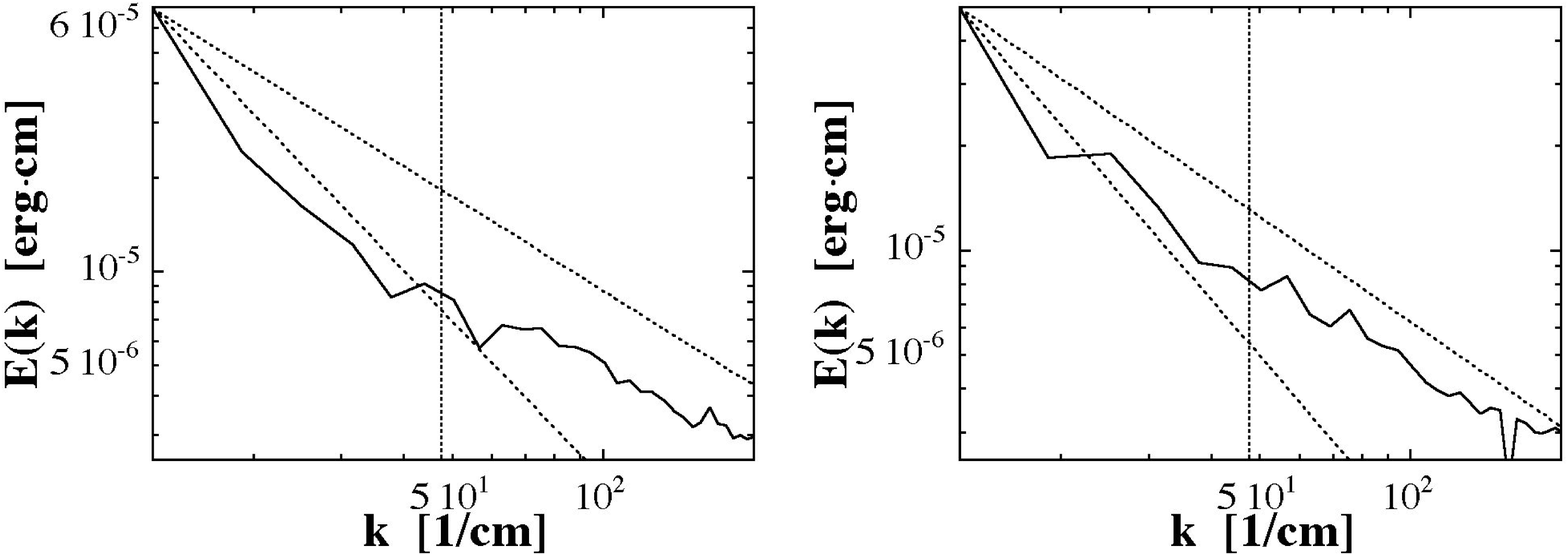}\\
 (c) \hspace{3cm} (d)
\end{center}
\end{minipage}
\caption{The energy spectra of the vortex tangle at $t$=0 sec(a), $t$=30.0 sec(b),
$t$=50.0 sec(c) and $t$=73.0 sec(d).}
 \label{eps4}
\end{figure}
 In Fig. 4(a) the spectrum may reflect an artifact of the initial configuration. However through the chaotic
dynamics, these spectra reach the equilibration(Fig. 4(d)). Figure 5 shows that the
time averaged spectrum of five configurations around $t=73.0$ sec.
\begin{figure}[tbhp]
\begin{center}
 \includegraphics[width=0.6\linewidth]{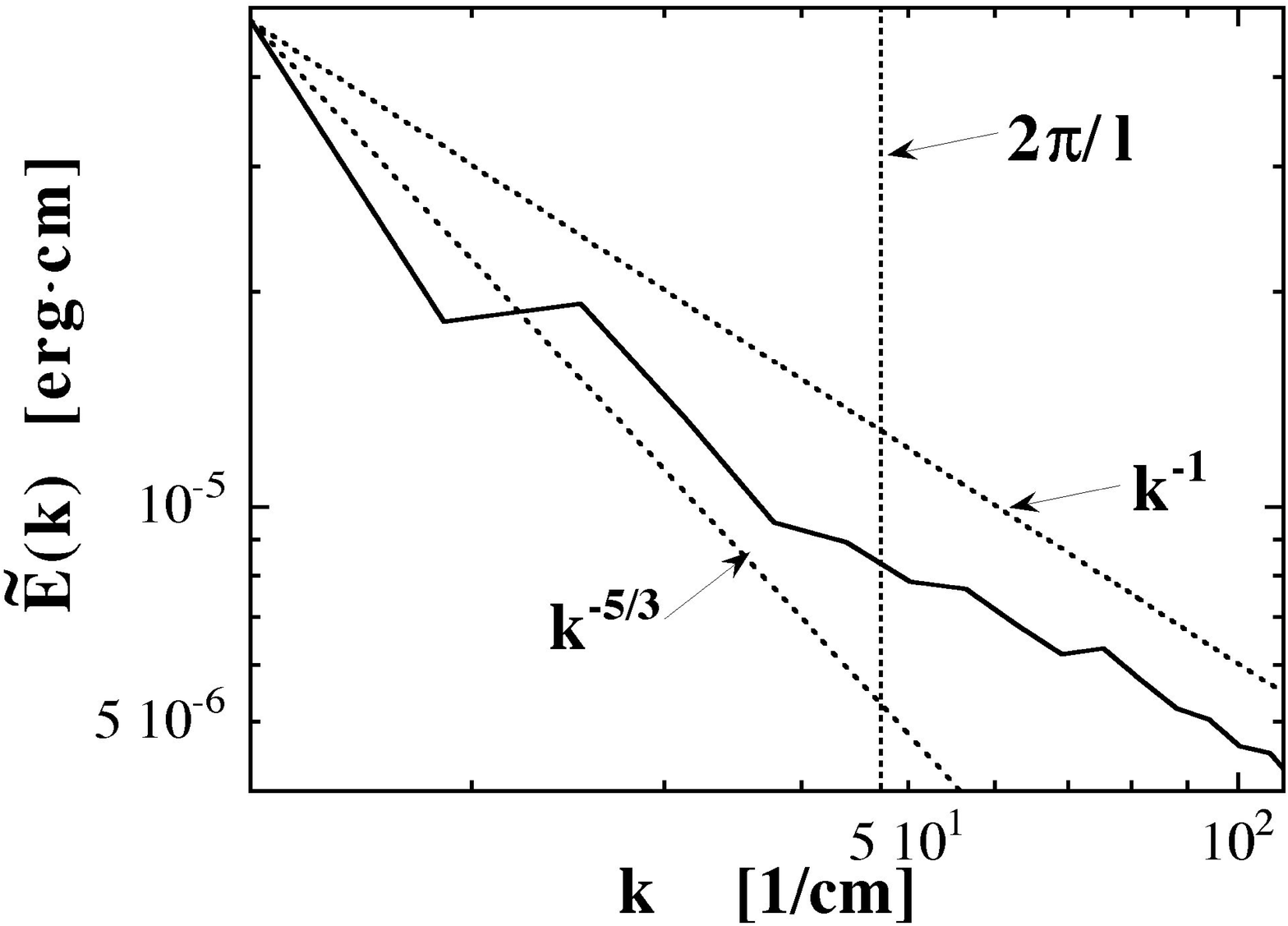}\\
\end{center}
\caption{The time averaged spectrum of five configurations around $t=73.0$ sec.}
 \label{eps5}
\end{figure}
 The slope is changed at $k \simeq 20$ cm$^{-1}$ and at $k=2\pi/l$, where $l$ is the intervortex spacing.
For $k>2\pi/l$, the energy spectrum has $k^{-1}$ behavior in the same manner as the
dilute vortex tangle, by the contribution of the isolated vortex lines. For $ 2\pi/L
< k < 20$ cm$^{-1}$, the spectrum is proportional to $k^{-5/2}$ by the effect of
boundary. The spectrum for 20 cm$^{-1}<k<2\pi/l$ is associated with velocity field
at scales larger than the intervortex spacing, and this slope is very similar to the
Kolmogorov law.

\section{CONCLUSIONS}

This work studies numerically the energy spectrum of the superfluid turbulence
without the mutual friction. For $k>2\pi/l$, the spectrum can be attributed to the
contribution of the individual vortex lines. For $2\pi/L<k<20$ cm$^{-1}$, the
spectrum depends on the boundary. In the intermediate range 20 cm$^{-1}<k<2\pi/l$,
the slope of the spectrum is very similar to the Kolmogorov law. The calculation of
the spectrum of a denser tangle is in progress so that the intermediate range may be
extended.

\end{document}